\documentclass[aps,prl,twocolumn,preprintnumbers,nofootinbib]{revtex4}
\usepackage{graphicx}

\begin{document}
\preprint{HUTP-02/A053}
\title{Light-sheets and Bekenstein's bound}
\author{Raphael Bousso}\email{bousso@physics.harvard.edu}
\altaffiliation{on leave from the University of California.}
\affiliation{%
Harvard University, 
Department of Physics, Jefferson Laboratory,
17 Oxford Street, Cambridge, MA 02138, U.S.A.; \\
Radcliffe Institute for Advanced Study, Putnam House, 
10 Garden Street, Cambridge, MA 02138, U.S.A.}%
\begin{abstract}
From the covariant bound on the entropy of partial light-sheets, we
derive a version of Bekenstein's bound: $S/M\leq \pi x/\hbar$, where
$S$, $M$, and $x$ are the entropy, total mass, and width of any
isolated, weakly gravitating system.  Because $x$ can be measured
along any spatial direction, the bound becomes unexpectedly tight in
thin systems.  Our result completes the identification of older
entropy bounds as special cases of the covariant bound.  Thus,
light-sheets exhibit a connection between information and geometry far
more general, but in no respect weaker, than that initially revealed
by black hole thermodynamics.
\end{abstract}

\pacs{03.67.-a, 04.70.Dy, 95.30.Sf}
                    
\maketitle

Entropy bounds have undergone a remarkable transformation from a
corollary to a candidate for a first principle~\cite{Bou02}.  After
proposing the generalized second law of thermodynamics
(GSL)~\cite{Bek72,Bek73}---that the sum of black hole entropy and
ordinary matter entropy never decreases---Bekenstein argued that its
validity necessitates a model-independent bound~\cite{Bek74,Bek81} on
the entropy $S$ of weakly gravitating systems:
\begin{equation}
S \leq \pi Md/\hbar.
\label{eq-bek1}
\end{equation}
where $M$ is the total gravitating energy, and $d$ is the linear size
of the system, defined to be the diameter of the smallest sphere that
fits around the system.  This inequality is obtained by considering
the classical absorption of the system by a large black hole; it does
not depend on the dimension of spacetime~\cite{Bou00b}.  Bekenstein's
bound is remarkably tight (consider, for example, a massive particle
in a box the size of its Compton wavelength).  It has appeared in
discussions ranging from information technology to quantum gravity.
Since $M \ll d/4G$ for a weakly gravitating system, it also implies
the ``spherical entropy bound'',
\begin{equation}
S \leq A_{\rm cs}/ 4G\hbar.
\label{eq-susskind}
\end{equation}
Here $A_{\rm cs}$ is the area of the circumscribing sphere.

Though confined to weak gravity, 't~Hooft~\cite{Tho93} and
Susskind~\cite{Sus95} ascribed fundamental significance to
Eq.~(\ref{eq-susskind}), claiming that it reflects a non-extensivity
of the number of degrees of freedom in nature.  This eventually
prompted the conjecture of a more general bound, the covariant entropy
bound~\cite{Bou99b}.  Empirically, this bound has been found to hold
in large classes of examples, including systems in which gravity is
the dominant force.  Meanwhile, no violation has been observed, nor
have any theoretical counterexamples been constructed from a realistic
effective theory of matter and gravitation.

Although the covariant bound does not conflict with the phenomenology
of our present models, it cannot be derived from known principles.  It
may be interpreted as an unexplained pattern in nature, betraying a
fundamental relation between information and spacetime geometry.  Then
the bound must eventually be explained by a unified theory of gravity,
matter, and quantum mechanics.  In the mean time, it should be
regarded as providing important hints about such a theory.

We are thus motivated to consider the covariant entropy bound primary,
and to try to derive other laws of physics from it.  As we will
shortly discuss, the bound has already been shown to imply the GSL, as
well as older, more specialized entropy bounds.  However, the oldest
(and, for weakly gravitating systems, tightest) bound of all,
Bekenstein's bound, is an exception.  It has not previously been
identified as a special case of the covariant bound.

The main purpose of this note is to fill this gap.  We will use the
covariant bound, in the stronger form of Ref.~\cite{FMW}, to derive an
inequality of the type introduced by Bekenstein, Eq.~(\ref{eq-bek1}).
Our result will be obtained directly, without use of the GSL.  Thus we
circumvent the continued debate of whether Bekenstein's bound is
really necessary for the GSL~\cite{Bek99,Wal99}.  

Let us briefly review the covariant bound, and its logical relation
to the GSL and to the spherical bound, Eq.~(\ref{eq-susskind}).  Given
any open or closed spatial surface $B$ at a fixed instant of time, one
can always construct at least two light-sheets.  A light-sheet of $B$
is a null hypersurface generated by non-expanding light-rays (i.e.,
null geodesics) which emanate from $B$ orthogonally~\cite{Bou99b}.
For example, for a spherical surface in Minkowski space, the two
light-sheets will be the two light-cones ending on $B$.

The covariant entropy bound~\cite{Bou99b} claims that the entropy $S$
of the matter on any light-sheet $L$ of $B$ is bounded by the surface
area $A(B)$:
\begin{equation}
S[L(B)]\leq A(B)/4G\hbar,
\label{eq-ceb1}
\end{equation}
where $G$ is Newton's constant.  (We set Boltzmann's constant and the
speed of light to 1.)  The entropy $S$ refers to the total
entropy of all matter systems that are ``seen'' by the light-rays
generating $L$ (systems whose worldvolume is fully intersected by
$L$).  

Let $B$ be a complete cross-section of the horizon of a black hole.
Then its past-directed ingoing light-sheet intersects with all the
matter systems that collapsed to form the black hole~\cite{FMW}.
Moreover, $A(B)/4G\hbar$ in this case represents the
Bekenstein-Hawking entropy of the black hole.  The bound thus
guarantees that the black hole entropy exceeds the matter entropy lost
to an outside observer.  That is, the GSL is upheld when a black hole
forms.

The GSL should also hold when a matter system falls into an existing
black hole.  In that case it requires that the black hole horizon area
increases enough so that the additional Bekenstein-Hawking entropy
compensates for the loss of matter entropy: $S \leq {\Delta A_{\rm
horizon} / 4 G\hbar}$.  In the form of Eq.~(\ref{eq-ceb1}), the
covariant entropy bound does not imply this relation.  This prompted
Flanagan, Marolf, and Wald~\cite{FMW} to propose a stronger
formulation, the ``generalized'' covariant entropy bound
(GCEB),
\begin{equation}
S[L(B;B')]\leq \frac{A(B)-A'(B')}{4G\hbar}.
\label{eq-ceb2}
\end{equation}
Here $A'$ is the area of any cross-sectional surface $B'$ on the
light-sheet $L$ of $B$.  $S$ denotes the entropy of matter systems
found on the portion of $L$ between $B$ and $B'$.\footnote{The GCEB
has been proven from a set of phenomenologically motivated
assumptions~\cite{FMW}, eliminating many possible counterexamples.
These assumptions, however, are not fundamental.  Unlike the GCEB
itself, they require the treatment of entropy as a fluid, which cannot
be appropriate at all scales.  Moreover, their plausibility is tied to
the thermodynamic limit.  To avoid confusion, we stress that we will
be assuming the GCEB axiomatically.  The assumptions and theorems of
Ref.~\cite{FMW} do not play any role here.}

Put differently, in constructing $L$, we are at liberty to follow each
light-ray until it intersects with neighboring light-rays.  (At these
caustic points the light-rays begin to diverge, and the non-expansion
condition becomes violated.)  But nothing forces us to follow each
light-ray to the bitter end.  We may construct a {\em partial
light-sheet\/} by terminating $L$ before caustics are reached.  Then
the endpoints of the light-rays will span a non-zero area $A'$.  It is
natural to expect that the inequality (\ref{eq-ceb1}) can be tightened
in this case, because we are not including in $S$ all the matter
systems that could have been reached by the light-rays.
Eq.~(\ref{eq-ceb2}) improves the bound accordingly. 

The GCEB does imply the GSL for all processes involving black holes,
including the absorption of a matter system by an existing black
hole~\cite{FMW}.  It also, of course, implies the weaker form of the
covariant bound (\ref{eq-ceb1}), which in turn implies the spherical
entropy bound (\ref{eq-susskind}) in weakly gravitating
regions~\cite{Bou99b}.

To derive Bekenstein's bound from the GCEB, we wish to apply
Eq.~(\ref{eq-ceb2}) to an isolated, weakly gravitating matter system.
The basic idea of our proof is to ``X-ray'' the system.  Because
matter bends light, initially parallel geodesics will arrive on the
``image plate'' slightly contracted.  The resulting area difference,
which bounds the system's entropy, will be expressed as the product of
the mass and the width of the system.

We make the following {\em assumptions}: ({\em i}\/)~The stress tensor
$T_{ab}$ has support only in a spatially compact region, the world
volume $W$ of the matter system.  ({\em ii\/})~Gravity is weak.
Specifically: ({\em ii\/}.1)~The metric is approximately flat: $g_{ab} =
\eta_{ab}+\delta g_{ab}$, with $\eta_{ab}=\mbox{diag}(-1,1,1,1)$ and
$|\delta g_{ab}|\ll 1$. ({\em ii\/}.2)~Any part of system is much
smaller than any (averaged) curvature scale it produces: $\langle
R_{abcd}\rangle_\ell\ll\ell^{-2}$, where $\langle
R_{abcd}\rangle_\ell$ is the average Riemann tensor along a distance
$\ell$.---It is believed that all physical matter (at least when
suitably averaged) satisfies the null and causal energy conditions.
These conditions may also be needed for the validity of the GCEB,
which however is being assumed here in any case.  To derive
Bekenstein's bound we shall require only the null energy condition:
({\em iii}\/)~$T_{ab} k^a k^b\geq 0$ for any null vector $k^a$.

\begin{figure}
\includegraphics[width=5cm]{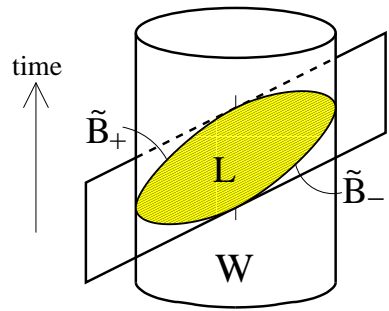}
\caption{\label{fig-bek} Matter system $W$, light-sheet $L$, entry
surface $\tilde{B}_+$, and exit surface $\tilde{B}_-$.  At first order
in $\delta g$, the bending of light leads to a small area difference between
entry and exit surfaces, which bounds the entropy of $W$.}
\end{figure}

We begin with some definitions valid at zeroth order in $\delta g$.
Cartesian coordinates $x^\mu$ $(\mu=0,\ldots,D-1)$ cover the
spacetime.  The corresponding vector fields $\partial/\partial x^\mu$
define an orthonormal frame at every point, which we take to be a rest
frame of $W$ for convenience.  (The remaining choice of spatial
orientation will be exploited later.)  The curves
\begin{equation}
x^0= x^1;~~~ (x^2,\ldots,x^{D-1})~\mbox{arbitrary constants},
\label{eq-l0}
\end{equation}
describe a set of parallel light-rays traveling in the $x^1$
direction (see Fig.~\ref{fig-bek}).
More precisely, they define a null geodesic congruence
$L$, with affine parameter $x^1$ and everywhere vanishing expansion.
We will be interested only in the intersection of the hypersurface $L$
with the world volume $W$ of the matter system.  Let $\tilde{B}_+$
$(\tilde{B}_-)$ be the set of the first (last) points of each
light-ray in $W$.  They form $(D-2)$-dimensional spatial surfaces
characterized by functions $x^1_\pm(x^2,\ldots,x^{D-1})$, with finite
range for $(x^2,\ldots,x^{D-1})$.  
(Connectedness is not necessary for this proof.)  All spatial sections
of a light-sheet are surface-orthogonal to the generating light-rays.
Hence, $L\cap W$ is a partial light-sheet with initial and final
surfaces $\tilde{B}_\pm$.  At zeroth order they have equal area.

In the exact metric, we may use the same coordinates. Generically,
however, the hypersurface $L$ as defined by Eq.~(\ref{eq-l0}) will be
neither null nor made of geodesics; nor is there a sense of strictly
non-positive expansion.  All of these qualitative conditions must hold
for $L$ to be a light-sheet; otherwise the GCEB cannot be applied.
Hence we must adjust $L$ slightly.  We will define two light-sheets,
$L_\pm$, both of which limit to $L$ as $\delta g \to 0$.

Consider the future- and $W$-directed light-rays orthogonal to
$\tilde{B}_+$.  Because gravity is weak, their expansion will be very
small (compared to the inverse width of $W$).  But it will not vanish
exactly, and it need not be of definite sign.  However, $\tilde{B}_+$
is embedded with codimension 1 in the boundary of $W$, $\partial W$.
Thus there exists a small (non-unique) deformation of $\tilde{B}_+$
within $\partial W$, the surface ${B_+}$, whose orthogonal null
geodesics have initially vanishing expansion {\em to all
orders}.\footnote{$B_+$ is constructed by matching the trace of its
extrinsic curvature in $\partial W$ to that of $\partial W$ in the
spacetime.  This prescription is not overconstrained, as it is
analogous to the construction of a minimal surface.}  Assumption ({\em
iii\/}) ensures that $\theta$ will not increase away from
$\tilde{B}_+$~\cite{Wald}, and ({\em ii\/}.2) excludes the possibility
that the light-rays intersect within $W$.\footnote{Strictly, this is
demonstrated by Eq.~(\ref{eq-decr}) below.}  Hence, the light-rays
generate a light-sheet $L_+$ that captures all of the matter system.
Let $A_+$ be the area of $B_+$, and let $A'_+$ be the area of the
surface $B'_+$ spanned by the same light-rays when they (last) leave
the system $W$.

Similarly, $\tilde{B}_-$ can be deformed to a surface ${B_-}$ of
exactly vanishing expansion.  This defines a second, slightly
different light-sheet $L_-$ with initial and final areas $A_-$ and
$A'_-$.  The light-sheets $L_\pm$ have opposite directions of
contraction, roughly $\pm x^1$.  We will be interested in the total
change of the cross-sectional area as each light-sheet traverses $W$:
$\Delta A_+\equiv A_+-A'_+$ and $\Delta A_-\equiv A_- -A'_-$.

Let $S$ be the entropy of the matter system, i.e., the logarithm of
the number of independent quantum states accessible to any system of
total mass $M$ occupying the world volume $W$ in a neighborhood of
$L_\pm$.  As each light-sheet fully contains the matter system, the
GCEB implies that $S\leq {\Delta A_+/ 4G\hbar}$ and also that $S\leq
{\Delta A_-/ 4G\hbar}$.  Therefore,
\begin{equation}
S\leq {\Delta A_+ + \Delta A_- \over 8G\hbar}.
\label{eq-doublefmw}
\end{equation}

To calculate $\Delta A_\pm$ to leading order, we may continue using
$(x^2,\ldots,x^{D-1})$ to label the light-rays in $L_\pm$.  We may
approximate the affine parameter along each ray by $\pm x^1$, and the
vector field tangent to the light-rays by
\begin{equation}
\pm k^a = \left({\partial\over\partial x^0}
+{\partial\over\partial x^1}\right)^a.
\label{eq-k}
\end{equation}

Let ${\cal A}_\pm$ be the cross-sectional area spanned by the
light-ray $(x^2,\ldots,x^{D-1})$ and its infinitesimally neighboring
light-rays in the light-sheet $L_\pm$, at the (affine) position $x^1$.
At each point on each of the two light-sheets, the expansion
$\theta_\pm$ is given by the trace of the null extrinsic
curvature~\cite{Bou02}.  Equivalently, it is the logarithmic
derivative of ${\cal A}_\pm$ with respect to the affine parameter $\pm
x^1$:
\begin{equation}
\theta_\pm(x^1;x^2,\ldots,x^{D-1}) = \pm \frac{d{\cal
A_\pm}/dx^1}{\cal A_\pm}.
\label{eq-thetadef}
\end{equation}

Raychaudhuri's equation,
\begin{equation}
{d\theta\over d(\pm x^1)} = - {1\over 2} \theta_\pm^2 -
\sigma_\pm^2 - R_{ab} k^a k^b,
\label{eq-raych}
\end{equation}
describes how the expansion changes along a light-ray.  Here $R_{ab}$
is the Ricci tensor.  There is no twist term because the light-sheets
are surface-orthogonal~\cite{Wald}.  The expansion and shear terms,
$\theta^2$ and $\sigma^2$, are of higher order than the stress term
and can be neglected.  In this approximation one can integrate
Eqs.~(\ref{eq-thetadef}) and (\ref{eq-raych}) for each light-ray:
\begin{eqnarray}
{\cal A_+}(x^1) 
& = & {\cal A}_+(x^1_+) 
\exp\int_{x^1_+}^{x^1} d\hat{x}^1 \theta_+(\hat{x}^1) \\
& = & {\cal A}_+(x^1_+) 
\left[ 1- \int_{x^1_+}^{x^1} d\hat{x}^1 
\int_{x^1_+}^{\hat{x}^1} d\hat{\hat{x}}^1
R_{ab} k^a k^b \right].
\label{eq-decr}
\end{eqnarray}
For $x^1=x^1_-$, the curvature term yields the fractional change in
each area element $dx^2\ldots dx^{D-1}$.  By assumption ({\em ii}.2),
this term will be small compared to unity.  The area change can be
integrated to obtain
\begin{equation}
\Delta A_+ = 8\pi G \int dx^2\ldots dx^{D-1} 
\int_{x^1_+}^{x^1_-} d\hat{x}^1
\int_{x^1_+}^{\hat{x}^1} d\hat{\hat{x}}^1 T_{ab} k^a k^b.
\end{equation}
After adding the analogous expression for the light-sheet $L_-$, the
integrals factorize, and Eq.~(\ref{eq-doublefmw}) becomes
\begin{equation}
S \leq {\pi\over\hbar} 
\int dx^2\ldots dx^{D-1}\, \Delta x^1
\int_{x^1_+}^{x^1_-} dx^1  T_{ab} k^a k^b.
\end{equation}

To continue the inequality, we replace the local width of the system,
$\Delta x^1\equiv x^1_--x^1_+$, by its largest value over
$(x^2,\ldots,x^{D-1})$, $x$.  (For convex systems, $x$ is the
separation of two planes orthogonal to $x^1$, which ``clamp'' $W$; but
generally, it can be smaller than that.)  This yields
\begin{equation}
S \leq \pi P_b k^b x/\hbar,
\end{equation}
where $P_b \equiv \int dx^1\ldots dx^{D-1}\, T_{ab} k^a$.  Note that
$P_b$ is a correctly normalized integral of the conserved tensor
$T_{ab}$ over a null hypersurface [see Eq.~(\ref{eq-k}) and, e.g.,
Appendix B.2 in Ref.~\cite{Wald}].  Since $T_{ab}$ vanishes outside
$W$, the hypersurface of integration can be extended to spatial
infinity without affecting the value of $P_b$.  Hence the time
component of $P_b$ is the total energy, and the (negative) spatial
components are the ADM momenta.  In a rest frame, the momenta vanish
by definition, and $P_0$ is equal to the system's total (``rest'')
mass $M$.  We thus obtain a ``generalized Bekenstein bound'',
\begin{equation}
S \leq \pi M x/\hbar.
\label{eq-bek2}
\end{equation}

Our result is somewhat stronger than the original Bekenstein bound,
Eq.~(\ref{eq-bek1}), because of our definition of the relevant length
scale, $x$.  Bekenstein advocated using the largest scale of the
system, the circumferential diameter $d$.  Our argument, however,
allows us to use the smallest dimension.  For example, if the system
is rectangular with sides of length $a<b<c$, we are free to align the
$x^1$ axis with the shortest edge, so that $x=a$.  For more general
shapes, the strength of Eq.~(\ref{eq-bek2}) is optimized as follows.
Find the greatest width of the system, $x(\Omega)$, for every
orientation $\Omega$ of the system relative to the $x^1$-axis; then
choose the particular orientation $\Omega_{\rm min}$ that yields the
smallest such greatest width, $x(\Omega_{\rm min})$.  If the shape of
the system is time-dependent, then $x$ can be minimized not only by
judicious rotations, but also by time-translations of $W$ relative to
$L$.\footnote{Obviously, boosts, rotations, and translations can
change the physical set-up only when applied either to $L$ or to $W$
alone.  Of these operations, only rotations and time-translations are
useful for minimizing the bound.  Spatial translations are either
trivial or equivalent to time translations.  Boosting $W$ is
equivalent to a rotation of $W$ followed by a boost in the $x^1$
direction.  The latter operation is actually trivial because $L$ is
invariant under such boosts.  Indeed, $\Delta x^1$ scales inversely
with $P_b k^b$ under $x^1$ boosts of $W$, so that one invariably
obtains the product of the rest frame quantities $x$ and $M$.}
Independently of the shape of the system, $x\leq d$ for all $\Omega$,
and in particular for $\Omega_{\rm min}$.  Hence Eq.~(\ref{eq-bek2})
implies Eq.~(\ref{eq-bek1}).  For systems with highly unequal
dimensions, such as a very flat box, $x\ll d$.  In this case
Eq.~(\ref{eq-bek2}) is much stronger than Eq.~(\ref{eq-bek1}).

The assumptions we stated earlier characterize the regime in which the
generalized Bekenstein bound can be applied.  Our construction will
not go through unless the system is compact and isolated, so that
initial and final surfaces of a suitable light-sheet can be
constructed.  The weakness of gravity ensures that the light-sheet
area decrease is small and that it is given by the product of a
(well-defined) width and mass.

Thus, our derivation does not give licence to all interpretations the
Bekenstein bound has received.  For example, we do not find support
for its application to a closed universe.  Let $S$ be the entropy of
the quantum fields on a spatial three-sphere of diameter $d$ at total
energy $M$.  (These quantities are well-defined in the absence of
gravity, $G=0$.)  In this case the system occupies a geometry which is
intrinsically curved.  Unlike an isolated system in flat space, it
cannot be fully covered by a partial light-sheet.  Hence, the
covariant bound does not imply Bekenstein's bound in this case.
Indeed, violations of Eq.~(\ref{eq-bek1}) were found for
supersymmetric conformal field theories on spatial spheres of various
dimensions~\cite{KutLar00}.

There is no evidence that the original Bekenstein bound is violated by
any complete, isolated, weakly gravitating system that can actually be
constructed in nature~\cite{Wal99,Bek00b}.  It also appears to be
reasonably tight, in that realistic matter can come within an order of
magnitude of saturating the bound~\cite{Bek81}.  But the generalized
Bekenstein bound faces challenges to which the original was immune.
Testing Eq.~(\ref{eq-bek2}) will important both in its own right, and
as a simple check of the GCEB that obviates the computation of
geodesics.  Detailed examples will be presented elsewhere.

We close on a speculative note.  Gravity plays a central role in our
derivation of Bekenstein's bound.  We combined the GCEB, a conjecture
involving the Planck area $G\hbar$, with classical equations involving
$G$.  But in due course $G$ dropped out, leaving only $\hbar$ in the
final result!  Indeed, Bekenstein's bound can be tested entirely
within quantum field theory, apparently without any use of the laws of
gravity~\cite{Bek81}.  This remarkable fact suggests a novel
perspective on the connection between gravity and quantum mechanics.
Note that for systems with small numbers of quanta ($S\approx 1$),
Bekenstein's bound can be seen to {\em require\/} non-vanishing
commutators between conjugate variables, as they prevent $Mx$ from
becoming much smaller than $\hbar$.  One is tempted to propose that at
least one of the principles of quantum mechanics implicitly used in
any verification of Bekenstein's bound will ultimately be recognized
as a consequence of Bekenstein's bound, and thus of the covariant
entropy bound and of the holographic relation it establishes between
information and geometry.

\acknowledgments

I would like to thank M.~Aganagic, J.~Bekenstein, \'E.~Flanagan, and
D.~Marolf for discussions.  I am especially grateful to H.~Casini for
stressing to me that flat systems may well obey Eq.~(\ref{eq-bek2}) if
boundary effects are properly taken into account.

\bibliographystyle{board}
\bibliography{all}

\end{document}